\documentclass{iopconfser}
\usepackage{notoccite}
\usepackage[num]{abntex2cite}				
\usepackage[num]{abntex2cite}
\usepackage[utf8]{inputenc}
\usepackage[utf8]{inputenc}
\usepackage[utf8]{inputenc}
\usepackage[english]{babel}
\usepackage{amsfonts}
\usepackage{amssymb}
\usepackage{verbatim}
\usepackage{xcolor}
\usepackage[T1]{fontenc}
\usepackage{graphicx}
\usepackage{physics}
\usepackage{siunitx}
\usepackage{cite}
\usepackage{makeidx}
\usepackage{helvet}
\usepackage{graphicx}
\usepackage{sectsty}
\usepackage{mdwlist}
\usepackage{tocbibind}
\usepackage{mathtools}
\usepackage{indentfirst}
\newcommand{\bea}   {\begin{eqnarray}}
\newcommand{\eea}   {\end{eqnarray}}
\newcommand{\cc}{”}
\def\zzg{${\mathbb Z}_2\times{\mathbb Z}_2$-graded }
\def\zgg{${\mathbb Z}_2\times{\mathbb Z}_2$ }
\begin{document}

\title{On braid statistics versus parastatistics}

\author{Francesco Toppan}

\affil{COTEO, Centro Brasileiro de Pesquisas Físicas, Rio de Janeiro (RJ), Brazil.}

\email{toppan@cbpf.br}

\begin{abstract}
I report the recent advances in applying (graded) Hopf algebras with braided tensor product in two scenarios:
{\it i}) paraparticles beyond bosons and fermions living in any space dimensions and transforming under the permutation group;
{\it ii}) physical models of anyons living in two space-dimensions and transforming under the braid group.  In the first scenario simple toy models based on the so-called $2$-bit parastatistics show that, in the multiparticle sector, certain observables can discriminate paraparticles from ordinary bosons/fermions (thus, providing a counterexample to the widespread belief of the ``conventionality of parastatistics\cc~argument).
In the second scenario the notion of (braided) Majorana qubit is introduced as the simplest building block to implement the Kitaev's proposal of a topological quantum computer which protects from decoherence.
\end{abstract}

\section{Introduction}

Paraparticles obeying a statistics beyond bosons/fermions were introduced by the Italian physicist Gentile in 1940 \cite{gen}; his idea was to extend the fermion statistics by accommodating at most $n$ paraparticles in a given quantum number (for fermions $n=1$). A consistent framework to introduce parastatistics, based on the so-called trilinear relations, was proposed in 1953 by Green \cite{gre}, see also \cite{grme}. The parastatistics whose paraparticles are exchanged via the permutation group (just like the ordinary bosons and fermions whose multiparticle wave functions are, respectively, totally symmetrized and totally antisymmetrized) have been widely investigated ever since.\par
Another breakthrough happened in the 1970's with the discovery \cite{lemy} that emergent particles constrained to live in two space dimensions can be exchanged via the most general braid group; for the permutation group an exchange operator $S_{1\leftrightarrow 2}$ squares to the Identity, so that $S_{1\leftrightarrow 2}^2={\mathbb I}$, while this is not necessarily the case for the braid exchange operator $B_{1\leftrightarrow 2}$ which admits the inequality $B_{1\leftrightarrow 2}^2\neq{\mathbb I}$ (the mathematical reason for this special properties of $2$ space dimensions is that the first homotopy group of the punctured plane is nonvanishing: $\pi_1({\mathbb R}^2_\ast)=\pi_1({\mathbb R}^2\backslash \{O\})={\mathbb Z}$).\par
Particles obeying the braid statistics have been named ``anyons\cc ~by Wilczek \cite{wil} (for a historical account on the prediction of anyons one can see \cite{gol} and the references therein). Quite recently, anyons have been experimentally detected (see the 2020 paper \cite{expanyons}).
Concerning anyons, an important suggestion was made by Kitaev (see \cite{kit}, \cite{brki} and \cite{nssfds}) to use emergent braided Majorana particles to encode the properties of  topological quantum computation. Unlike ``ordinary\cc ~quantum computers which manipulate qubits, the proposed topological quantum computer should manipulate braided Majorana fermions which offer topological protection under the quantum decoherence.\par
Concerning paraparticles exchanging under the permutation group the situation is different. Soon after the introduction of the consistent \cite{gre} Green's parastatistics physicists started wondering why these paraparticles 
were not observed (unlike bosons and fermions) in Nature. Some sort of consensus started to emerge: maybe these paraparticles cannot be experimentally detected because their manifestations can always be reconstructed via ordinary bosons/fermions?  A mathematical result by Araki \cite{ara} seemed to point into this direction:
parafields can be reconstructed from ordinary fields by applying Klein operators. Even so, this property cannot dismiss the existence of observable paraparticles (an analogous property is applied to low-dimensional fermions which can be reconstructed, through ``bosonization\cc , from nonlocal bosonic operators \cite{col}). 
Probably the most complete reference discussing several proposed arguments to justify the ``conventionality of parastatistics\cc ~is  \cite{conventionality}. As recalled in that work,  none of the supporting arguments is fully satisfactory. The most stringent one is perhaps the
Doplicher-Roberts Reconstruction Theorem \cite{doro} which is based on two underlying hypotheses: the existence of a superselection rule and the presence of a localization principle. Unfortunately, a widespread misunderstanding of these arguments which were extrapolated beyond their realm of validity generated prejudices which seriously hampered, as it will be discussed in Section {\bf 3}, the development of certain theoretical disciplines. The first counterexamples to the conventionality argument (based on the parastatistics induced by the Rittenberg-Wyler\cite{{rw1},{rw2}}  ${\mathbb Z}_2^2$-graded color Lie (super)algebras) have been presented in \cite{top1} and \cite{top2}; it was shown that certain eigenvalues measured by observables in the multiparticle sectors of  paraparticle Hamiltonians could not be recovered from ordinary boson/fermion statistics. In \cite{{top1},{top2}} the parastatistics were introduced through the First Quantization framework proposed in \cite{maj} and based on a
(graded) Hopf algebra endowed with a braided tensor product. \par
The same Hopf algebra framework with braided tensor product was later applied in \cite{top3} to define the braiding of a certain class of Majorana fermions which can implement the Kitaev's proposal. The idea behind \cite{top3} was to introduce the simplest possible braiding system. Ordinary computers manipulate Boolean logic based on bits of information; quantum computers manipulate entangled qubits; the proposed topological quantum computer should manipulate braided Majorana fermions (that is, fermions which coincide with their own antiparticle). The simplest of such settings consists of ${\mathbb Z}_2$-graded qubits (the ``Majorana qubits\cc ) which can be braided via 
the $R$-matrix of the Alexander-Conway polynomial in the linear crystal rep on exterior algebra introduced in \cite{kasa}.\par
~\par
The scheme of the paper is the following. In Section {\bf 2} the First Quantization framework based on the graded Hopf algebra endowed with a braided tensor product is briefly recalled. Section {\bf 3} describes the present state of the art concerning the theoretical detectability (and the experimental challenges) of paraparticles which  are exchanged via the permutation group. Section {\bf 4} introduces the braided Majorana qubits, showing that they implement a Gentile-type parastatistics. The very recent \cite{top4} results are reported: the truncations of the multiparticle spectra are derived from a quantum group representation.  Furthermore, the notion of ``metasymmetry\cc ~defined in \cite{lese1} is shown to be applied. It leads to a mixed-bracket generalization of the Heisenberg-Lie algebras, the mixed brackets being interpolation of the ordinary commutators and anticommutators.

\section{Graded Hopf algebras endowed with a braided tensor product}

The \cite{maj} framework of a graded Hopf algebra endowed with a braided tensor product applies to the First Quantization (the construction of the multiparticle sectors) of a quantum Hamiltonian. The scope of this framework is limited because it
only deals with linear theories (nonlinear interactions have to be dealt with a different approach); it is nevertheless sufficient for several relevant applications including the ones here considered. It should also be mentioned that the class of integrable systems (which, essentially, can be regarded as linear systems in disguise) can be investigated within this framework.\par
The connection between the \cite{maj} approach to parastatistics and the traditional \cite{{gre},{grme}} framework based on the trilinear relations has been discussed in \cite{{anpo1},{kada1}}.\par
The starting point is a Universal Enveloping Algebra  ${\cal U}\equiv {\cal U}({\mathfrak{g}})$) of a graded Lie algebra
${\mathfrak g}$. ${\cal U}$ is a graded Hopf algebra possessing compatible structures (unit and multiplication), costructures (counit and coproduct) and antipode. Particularly relevant for our scopes is the coproduct $\Delta$, which is a map
\bea\label{coproduct}
\Delta &:&{\cal U}\rightarrow {\cal U} \otimes_{br} {\cal U}
\eea
satisfying the coassociativity property
\bea\label{coassociativity}
   (\Delta\otimes_{br} id)\Delta(U)&=&(id\otimes_{br} \Delta)\Delta(U) \qquad {\textrm{for}}\quad U\in {\cal U},\nonumber\\
\Delta^{(n+1)} &=& (\Delta\otimes_{br} id)\Delta^{(n)}=(id\otimes_{br} \Delta)\Delta^{(n)}.
\eea
For any $U_A,U_B\in {\cal U}$, the further property
\bea\label{uaubprop}
   \Delta(U_AU_B)&=&\Delta(U_A)\Delta(U_B)
\eea
implies that the action on any given $U\in {\cal U}({\mathfrak g})$ is recovered from the action of the coproduct on the
Hopf algebra unit ${\bf 1}$ and the Lie algebra elements $g\in {\mathfrak{g}}$; they are given by
\bea\label{deltaidg}
   \Delta({\bf 1})={\bf 1}\otimes_{br}{\bf 1}, &\quad & 
   \Delta({ g})={\bf 1}\otimes_{br}{g}+g\otimes_{br} {\bf 1}.
\eea
In the above formulas the convenient symbol $\otimes_{br}$ of the braided tensor product (which will be defined later) appears. 
Let $R$ be a representation of the Universal Enveloping Algebra ${\cal U}$ on a vector space $V$. The representation of the operators induced by the coproduct will be denoted with a hat:
\bea\label{rep}
&{\textrm{for}}\quad R: {\cal U}\rightarrow V,\qquad {\widehat \Delta}:= \Delta|_R\in End (V\otimes V),\qquad 
 {\textrm{with}}\quad  {\widehat{ \Delta(U)}}\in V\otimes V. &
\eea
It follows from the coassociativity property that
\bea
{\widehat{ \Delta^{(n)}(U)}}&\in& V\otimes \ldots \otimes V\qquad (n+1 \quad {\textrm{times}}).
\eea
One should note that, when considering a representation, the braided tensor product $\otimes_{br}$ is replaced by the ordinary tensor product $\otimes$.\par
This formalism allows to construct the braided $N$-particle Hilbert space ${\cal H}^{(N)}$ as a subset of the tensor products of $N$ single-particle Hilbert spaces ${\cal H}^{(1)}$. Let's set, for simplicity, ${\cal H}\equiv{\cal H}^{(1)}$. We have
\bea
{\cal H}^{(N)}&\subset &{\cal H}^{\otimes N}.
\eea
The $N$-particle vacuum state $|vac\rangle_{N}$ is the tensor product of $N$ single-particle vacua:
\bea
\qquad |vac\rangle_N &=& |vac\rangle \otimes \ldots \otimes |vac\rangle\qquad \qquad (\textrm{$N$ times}).
\eea
The $N$-particle Hamiltonians $H_{(N)}$ are obtained by applying the $N$-particle coproducts $\Delta^{(N-1)}$ to the single-particle Hamiltonian $H$:
\bea\label{ncoproduct}
H_{(N)} &=&{\widehat{\Delta^{(N-1)}(H)}}
\eea
(one should not confuse the calligraphic symbol ${\cal H}$ of the Hilbert space with the symbol $H$ which denotes a Hamiltonian).\par
The introduction of a nontrivial braiding requires specifying how the graded Lie algebra generators are braided in the tensor product. The braidings of the elements of the Universal Enveloping Algebra  are obtained, from (\ref{uaubprop}), as a consequence. Let $a,b,c,d$ be four generators of a graded Lie algebra. The braided tensor product can be expressed as
\bea
(a\otimes_{br} b)\cdot (c\otimes_{br} d) &=& (a\otimes_{br} {\bf 1})\cdot \Psi(b,c)\cdot ({\bf 1}\otimes_{br} d),\nonumber\\
({\bf 1}\otimes_{br} b)\cdot (c\otimes_{br}{\bf 1})&=& \Psi(b,c),
\eea
where the braiding operator $\Psi(b,c)$ satisfies \cite{maj} a set of braided consistency conditions. \par
In the simplest example, for a fermionic nilpotent creation operator $f^\dagger$ satisfying $(f^\dagger)^2=0$, the braiding is just recovered from a $-1$ sign:
\bea
({\bf 1}\otimes_{br} f^\dagger)  \cdot (f^\dagger\otimes_{br}{\bf 1})&=& -(f^\dagger\otimes_{br}f^\dagger).
\eea
It follows that the coproduct $\Delta(f^\dagger)$ is nilpotent:
\bea
(\Delta(f^\dagger))^2 &=&  (f^\dagger\otimes_{br}{\bf 1}+ {\bf 1}\otimes_{br} f^\dagger)  \cdot (f^\dagger\otimes_{br}{\bf 1}+ {\bf 1}\otimes_{br} f^\dagger) = \nonumber\\
&=& (f^\dagger\otimes_{br}{\bf 1})\cdot ( {\bf 1}\otimes_{br} f^\dagger)+
( {\bf 1}\otimes_{br} f^\dagger)\cdot (f^\dagger\otimes_{br}{\bf 1})=\nonumber\\&=& (f^\dagger\otimes_{br}f^\dagger)-(f^\dagger\otimes_{br}f^\dagger)=0.
\eea
The above nilpotency condition implies, as a physical interpretation, the Pauli exclusion principle of ordinary fermions.\par
In the following we will present examples of more complicated braided consistency conditions.

\section{Detectability of paraparticles: state of the art}

In this Section I will present the state of the art concerning the detectability of paraparticles which exchange positions under the permutation group, by introducing the recent theoretical (and experimental) advances. Before that, I discuss the simplest (minimal) example of parastatistics, the $2$-bit parastatistics induced by Rittenberg-Wyler \cite{{rw1},{rw2}} (and Scheunert\cite{sch}) color Lie (super)algebras.\par
~\par
Ordinary ($1$-bit) physics based on bosons/fermions is recovered from a ${\mathbb Z}_2$-graded Lie superalgebra,
while the two admissibile clases of $2$-bit parastatistics are recovered from two different \zzg  Lie (super)algebras:
\bea\label{threecases}
{\textrm{{\it ~~~ i})}} && {\textrm{${\mathbb Z}_2$-graded Lie superalgebras,}} \nonumber\\
{\textrm{{\it ~~ ii})}} &&{\textrm{\zzg Lie superalgebras and}}\nonumber\\
{\textrm{{\it ~ iii})}} &&{\textrm{\zzg Lie algebras. }}
\eea
Depending on the case under consideration, a graded Lie (super)algebra ${\mathfrak g}$ is decomposed into 
\bea\label{A1}
i)~~~~~&:&{\mathfrak g} = {\mathfrak g}_0\oplus {\mathfrak g}_1,\nonumber\\
ii) ~{\textrm{and}}~ iii)&:& {\mathfrak g} = {\mathfrak g}_{00}\oplus {\mathfrak g}_{01}\oplus {\mathfrak g}_{10}\oplus {\mathfrak g}_{11}.\eea
The even ($0$) and odd ($1$) generators in {\it i}) are bosonic (fermionic). The four sectors of {\it ii}) and {\it iii}) are described by $2$ bits. The grading of a generator in {\it i}) is given by ${\vec \alpha}\equiv \alpha\in\{0,1\}$. The grading of a generator in
{\it ii}) and {\it iii}) is given by the pair ${\vec \alpha}=(\alpha_1,\alpha_2)$, with $\alpha_{1,2}\in\{0,1\}$.\par
Three respective inner products, with addition ${\textrm{mod}}~2$, are defined as:
\bea\label{innerproducts}
~ {i})~: ~~ {\vec\alpha}\cdot{\vec\beta} &:=& \alpha\beta\in \{0,1\},\nonumber\\
~ {ii})~: ~~ {\vec\alpha}\cdot{\vec\beta} &:=& \alpha_1\beta_1+\alpha_2\beta_2\in \{0,1\},\nonumber\\
~ {iii})~:~~  {\vec\alpha}\cdot{\vec\beta} &:=& \alpha_1\beta_2-\alpha_2\beta_1\in \{0,1\}.
\eea
The graded algebra ${\mathfrak g}$ is endowed with the operation $(\cdot,\cdot):{\mathfrak{g}}\times{\mathfrak g}\rightarrow{\mathfrak g}$. \par
Let $a,b,c\in{\mathfrak{g}}$ be three generators whose respective gradings
are ${\vec{\alpha}}, {\vec{\beta}},{\vec{\gamma}}$. The bracket $(\cdot,\cdot)$, defined as
\bea\label{roundbracket}
(a,b)&:=& ab -(-1)^{{\vec\alpha}\cdot{\vec \beta}}ba,
\eea
results in either commutators or anticommutators.\par
The operation satisfies the graded Jacobi identities
\bea
\label{gradedjac}
 (-1)^{\vec{\gamma}\cdot\vec{\alpha}}(a,(b,c))+
 (-1)^{\vec{\alpha}\cdot\vec{\beta}}(b,(c,a))+
 (-1)^{\vec{\beta}\cdot\vec{\gamma}}(c,(a,b))&=&0.
\eea

The grading $\deg[(a,b)]$ of the generator $(a,b)$ is the ${\textrm{mod}}~2$ sum
\bea\label{gradingcomp}
\deg[(a,b)]&=& {\vec{\alpha}}+{\vec{\beta}}.
\eea

The construction of the multi-particle states is based, as discussed in the previous Section, on the notions of coproduct and  braided tensor product.  Extra signs, induced by the \zgg  gradings, enter the braided tensor product
with respect to the $1$-bit boson/fermion statistics. The following tables present the two inequivalent $2$-bit parastatistics:
\bea \label{2cases}
ii)\quad
\begin{array}{c|cccc}&00&10&01&11\\ \hline 00&0&0&0&0\\10&0&1&0&1\\01&0&0&1&1\\11&0&1&1&0
\end{array}, &\quad & iii)\quad 
\begin{array}{c|cccc}&00&10&01&11\\ \hline 00&0&0&0&0\\10&0&0&1&1\\01&0&1&0&1\\11&0&1&1&0
\end{array}. 
\eea
The corresponding paraparticles are accommodated into $4$ sectors. The entries $0$ and $1$ respectively denote
a commutator or an anticommutator. In the $ii$) superalgebra case the sectors $10$ and $01$ are fermionic in nature
due to the presence of $1$ in the diagonal entries (which implies a Pauli exclusion principle). On the other hand, unlike ordinary fermions, $10$ and $01$ particles mutually commute, producing a pair of \zzg parafermions. The $00$ sector is purely bosonic, while the consistency of the construction requires the introduction of a $11$ sector which has a bosonic nature (due to the $0$ entry in the diagonal), but which anticommutes with the parafermions. The alternative construction $iii$) produces one bosonic ($00$) and three parabosonic sectors $(10,01,11)$ which mutually anticommute.\par
The \zzg color Lie superalgebra $ii$ is a true generalization of ordinary physics. Indeed, it is immediate to see that, by leaving empty the $01$ and $11$ sectors, the $00$ and $10$ sectors alone reproduce the ordinary superalgebra of bosons/fermions. Being a true generalization one could have expected that the physical consequences of the $2$-bit parastatistics would have been systematically investigated from the very beginning.  This turned out not to be the case due to the prejudice that the parastatistics can be recovered from ordinary bosons/fermions statistics. Some physical applications were discussed, see \cite{vas}, but a systematic investigation was delayed for many years. The situation changed recently with several advances. It was recognized that \zzg color Lie superalgebras are dynamical symmetries of nonrelativistic L\'evy-Leblond spinors \cite{{aktt1},{aktt2},{ryan}}, classical \cite{{akt1},{brusigma}} and quantum \cite{{brdu},{akt2}} models were introduced, \zzg superspace was proposed \cite{{pon},{aido1},{aizt}}, bosonization approach presented \cite{que}, \zzg integrable systems introduced \cite{{bruSG},{z2z2int}}. \par
 Along the years \zzg extensions of parastatistics were investigated \cite{{yaji},{tol1},{stvj1},{stvj2}}, but the specific issue of the conventionality argument was not addressed. This question became pressing when Bruce-Duplij produced \cite{brdu} the matrix quantum Hamiltonian 
\bea\label{qham12100}
H&=&{\footnotesize{\frac{1}{2} \left(\begin{array}{cccc}-\partial_x^2+W^2+W'&0&0&0\\ 0&-\partial_x^2+W^2+W'&0&0\\ 0&0&-\partial_x^2+W^2-W'&0\\ 0&0&0&-\partial_x^2+W^2-W'\end{array}\right)}}
\eea
(for $W\equiv W(x)$ and $W'\equiv \frac{d}{dx}W(x)$), which is at the same time a supersymmetric quantum Hamiltonian and invariant under a \zzg one-dimensional super-Poincar\'e algebra. Does its \zzg invariance produce physical measurable consequences? In order to answer this question I had to simplify the problem, reducing it to a purely combinatorial problem for a matrix harmonic oscillator by setting $W(x)=x$. For the single-particle quantum Hamiltonian (\ref{qham12100}) the \zzg invariance is an alternative, physically equivalent, description as the one given by the supersymmetric quantum mechanics interpretation. The situation changes in the multiparticle sector
of the model, which is treated with the graded Hopf algebra formalism (endowed with the braided tensor product) as described in Section {\bf 2}. \par
 The construction goes as follows: a $2$-particle observable belonging to the $00$ (bosonic) sector of the theory applied to a certain state produces an eigenvalue which corresponds  to a $\pm 1$ sign. The sign is different if the state is constructed from a pair of ordinary fermions or from a pair of \zzg parafermions. 
The theories with \zzg parabosons, resulting from constructions based on the so-called minimal \zzg Lie algebras and superalgebras (see \cite{kuto}), are treated in a similar way and present analogous results \cite{top2}.

\subsection{The present theoretical and experimental situation}

The models analyzed in \cite{{top1},{top2}} allow to evade the Doplicher-Roberts \cite{doro}  conventionality of parastatistics argument because, being First-Quantized theories, do not admit a localization principle. The same is true for the First-Quantized models presented in \cite{nbits} (see also \cite{toptransm}). They are derived from all possible ``statistical transmutations\cc~ of the operators entering a de Alfaro-Fubini-Furlan \cite{dff} matrix deformed oscillator. For this class of models, the detectable effects of the \zzg parastatistics can be directly determined  by the degeneracy levels of certain multiparticle energy eigenvalues. Here as well, the localization principle plays no role.\par
A different mechanism to evade the localization principle is presented in \cite{waha}. The adopted basic strategy is the same as in \cite{{top1},{top2}}, i.e., showing that the parastatistics results cannot be reproduced by ordinary bosons/fermions. The class of models under consideration differs: spin chains of paraoscillators are introduced.
The Doplicher-Roberts argument is evaded because the excited states are created by non-localized operators which are of stringy nature.\par
These recent theoretical advances imply that the parastatistics based on the permutation group are in principle experimentally detectable. What about the present experimental situation? A very promising line is based on the results of \cite{parasim} (for a simulation of parastatistics) and \cite{paraexp} (for an experimental realization).
The main point to be stressed is that, in a laboratory experience, they provide a mechanism to manipulate the parastatistics; the key ingredient is a ion trap with two energy levels which can be modulated using laser pulses.\par
Does it mean that we have experimentally detected or produced in the laboratory paraparticles transforming under the permutation group? Well, the answer is subtle: not yet. The reason is that theoreticians and experimentalists
should meet halfway, so to speak. In performing an experiment one should be sure that it produces an inequivocal signature of a parastatistics. This would be the case if the experimental model under investigation is known to be described by
a theoretical model  which, on theoretical grounds, is known not to be reproducible by ordinary bosons/fermions. As discussed at length, examples of such models are the ones introduced in \cite{{top1},{top2},{waha},{nbits}}.\par
~\par
In a related area another possible scenario for the detection of \zzg paraparticles is condensed matter and, specifically, the topological superconductors and insulators. The {\it periodic table of topological superconductors and insulators}, based on the so-called $10$-fold way, is a well-known tool in the classification of these materials, see \cite{{kit2},{zir},{alzi},{rsfl}}. The $10$-fold way refers to the connection (see \cite{bae} for a  simple mathematical presentation) with the $10=3+7$ associative division and superdivision algebras. The $3$ associative division algebras are the well-known real numbers, complex numbers and quaternions.  $7$ more ${\mathbb Z}_2$-graded superdivision algebras have to added (producing the $10$-fold way). The elements of a ${\mathbb Z}_2$-graded superdivision algebra are split into even and odd sectors, where each sector is invertible. Furthermore, a superselection rule prevents introducing superpositions (linear combinations) of  generators of even and odd sectors.
The superselection prevents constructing divisors of zero. In \cite{aaca} we classified with Z. Kuznetsova the \zzg superdivision algebras over the reals, finding $13$ extra cases that have to be added to the $10$-fold way. We also pointed out that Hamiltonians constructed with parafermionic oscillators can be accommodated into the \zzg superdivision algebra setting (see \cite{aaca} for details).

\section{Braided Majorana qubits and related mathematical structures}

We review at first the multiparticle quantization of braided Majorana qubits following \cite{top3} and \cite{topscipost}. A single ${\mathbb Z}_2$-graded Majorana qubit defines a bosonic  vacuum state $|0\rangle$ and a fermionic excited state $ |1\rangle$:
{\small{\bea\label{qubit01}
|0\rangle = \left(\begin{array}{c} 1\\0\end{array}\right) , &\quad& |1\rangle =\left(\begin{array}{c} 0\\1\end{array}\right) .
\eea}}
The following operators, acting on the ${\mathbb Z}_2$-graded qubit, close the ${\mathfrak{gl}}(1|1)$ superalgebra:
{\small{\bea\label{4op}
&\alpha =\left(\begin{array}{cc} 1&0\\0&0\end{array}\right),\quad ~\beta =\left(\begin{array}{cc} 0&1\\0&0\end{array}\right),\quad ~ \gamma =\left(\begin{array}{cc} 0&0\\1&0\end{array}\right),\quad ~\delta =\left(\begin{array}{cc} 0&0\\0&1\end{array}\right).&
\eea}}
The diagonal operators $\alpha,\delta$ are even, while $\beta,\gamma$ are odd ($\gamma$ is the fermionic
creation operator). The presence of a real structure implies that the fermionic excited state coincides with its own antiparticle (i.e., it is Majorana). The diagonal operator $\delta$ can be assumed to be the single-particle Hamiltonian $H_1$ (therefore, $H_1\equiv \delta$).\par 
Bosons/fermions satisfy a superselection rule which implies that they cannot be superposed (linearly combined). It follows that the ``Bloch sphere\cc ~ of the  Majorana qubit, determined by a ray vector, instead of being $S^2$ as an ordinary qubit, is given by ${\mathbb Z}_2$, i.e. a one bit of information; the identifications are $0\equiv |0\rangle$, $1\equiv |1\rangle$. The nontrivial braiding which produces the braided multiparticle Majorana qubits is given by 
the $t$-dependent $4\times 4$ constant matrix $B_t$ (invertible for the complex parameter  $t\neq 0$),  given by
{\footnotesize{\bea\label{btmatrix}
B_t&=&\left(\begin{array}{cccc} 1&0&0&0\\0&1-t&t&0\\0&1&0&0\\0&0&0&-t\end{array}\right).
\eea
}} 
$B_t$ is the $R$-matrix of the Alexander-Conway polynomial in the linear crystal rep on exterior algebra, see
\cite{kasa}. $B_t$ defines the braiding of the  fermionic creation operator $\gamma $ according to
\bea \label{braidingamma}
({\mathbb I}_2\otimes_{br} \gamma)\cdot (\gamma\otimes_{br} {\mathbb I}_2) &=& B_t\cdot (\gamma\otimes_{br} {\mathbb I}_2)\cdot ({\mathbb I}_2\otimes_{br} \gamma) \equiv B_t\cdot (\gamma\otimes_{br}\gamma).
\eea
The braiding compatibility  condition is guaranteed by $B_t$ satisfying the braid relation
\bea\label{braidedrel}
(B_t\otimes {\mathbb I}_2)\cdot ({\mathbb I}_2\otimes B_t)\cdot 
(B_t\otimes {\mathbb I}_2) &=& ({\mathbb I}_2\otimes B_t) \cdot
(B_t\otimes {\mathbb I}_2)\cdot ({\mathbb I}_2\otimes B_t).
\eea
The Hopf algebra which, together with the braided tensor product $\otimes_{br}$, defines the multiparticle sectors is
the graded Universal Enveloping Algebra ${\cal U}\equiv {\cal U}({\mathfrak{gl}}(1|1))$. 

This framework induces the braided multiparticle sectors of the Majorana qubits which implement a Gentile-type parastatistics with at most $s$ particles accommodated into an $N$-particle sector, for the integer values $s=2,3,4,5,\ldots $. In a convenient parametrization, see \cite{top4},  the inequivalent physics only depends on the roots of unity values of $t$ given by
\bea\label{ts}
t_s &=& e^{\pi i (\frac{2}{s}-1)}.
\eea
For finite values of $s=2,3,4,\ldots$ we get a truncated spectrum of the $N$-particle energy eigenvalues:
\bea\label{truncatedenergy}
E &=& 0,1,\ldots, N\qquad \quad~{\textrm{for}}\quad N<s,\nonumber\\
E &=& 0,1,\ldots, s-1 \qquad {\textrm{for}} \quad N\geq s;
\eea
(a plateau is reached at the maximal energy level $s-1$; this is the maximal number of excited particles that can be accommodated in a multi-particle Hilbert space).\par
The $s\rightarrow\infty$ limit produces $t=-1$ and an untruncated spectrum for the $N$-particle energy eigenvalues $E$:
\bea\label{genericenergy}
E &=& 0,1,\ldots, N\qquad {\textrm{for any given}}\quad N;
\eea
in this case there is no plateau and the maximal energy eigenvalues grow linearly with $N$.\par

A first open question was answered in the recent published paper \cite{top4}. The truncated spectrum is reminiscent of the truncations at roots of unity \cite{{lus},{dck}} recovered from quantum group representations. On the other hand, the formalism of classical  ${\cal U}({\mathfrak{gl}}(1|1))$ Universal Enveloping Algebra supplemented by the braided tensor product $\otimes_{br}$ from (\ref{braidingamma}) does not directly encode quantum group data. \par
It turns out that this construction can be reproduced from the quantum group ${\cal U}_q({\mathfrak{osp}}(1|2))$, introduced in \cite{kure}, by selecting a special class of multiparticle representations. The truncations at roots of unity observed in the previous framework are directly related to the representations at $q$ roots of unity for ${\cal U}_q({\mathfrak{osp}}(1|2))$.\par
~\par
A different fascinating connection is discovered when the braided tensor product $\otimes_{br}$, instead of simply {\it declaring} to be braided, is realized as an ordinary $\otimes$ tensor product via the introduction of intertwining operators. As an example, the mappings 
\bea\label{matrixrepof}
 (\gamma~\otimes_{br}{\mathbb I}_2)\mapsto \gamma\otimes {\mathbb I}_2, &\qquad
 ({\mathbb I}_2~\otimes_{br}\gamma)\mapsto W_t\otimes \gamma.
\eea
allow to recover the $\otimes_{br}$ braiding relations 
\bea
 ({\mathbb I}_2~\otimes_{br}\gamma)\cdot 
 (\gamma~\otimes_{br}{\mathbb I}_2)&\mapsto& (W_t\otimes\gamma)\cdot(\gamma\otimes {\mathbb I}_2)= (W_t\gamma)\otimes\gamma,\nonumber\\
 (\gamma~\otimes_{br}{\mathbb I}_2)\cdot ({\mathbb I}_2~\otimes_{br}\gamma)&\mapsto&
(\gamma\otimes {\mathbb I}_2)\cdot (W_t \otimes\gamma) = (\gamma W_t)\otimes\gamma,
\eea
provided that the $2\times 2$ 
intertwining operator $W_t$ satisfies the consistency condition
\bea\label{consistencyintertwining}
W_t\gamma &=& (-t) \gamma W_t.
\eea
A solution, expressed in terms of the  $t_s=-e^{\frac{2i\pi}{s}}$ position, is 
\bea\label{solutionintertwining}
W_{t_s}&=& \cos(\frac{-\pi}{s} )\cdot {\mathbb I_2} + i \sin(\frac{-\pi}{s} )\cdot X,\qquad {\textrm{where $X=$ {\footnotesize  $\left(\begin{array}{cc} 1&0\\0&-1\end{array}\right)$.}}}
\eea
Indeed, we get
\bea
W_{t_s}\gamma &=& e^{\frac{2\pi i}{s} }\gamma W_{t_s}.
\eea
This allows to make a connection with the notion of ``symmetries wider than supersymmetry\cc~ presented by Leites and Serganova in \cite{lese1} (see also \cite{lese2}); this notion concerns the existence of statistics-changing maps which do not preserve  the ${\mathbb Z}_2$-grading of ordinary Lie superalgebras.\par
Leites-Serganova introduced the concepts of ``metamanifolds\cc~(as an extension of supermanifolds),
``metaspace\cc~(as an extension of superspace) and ``metasymmetry\cc~(as an extension of supersymmetry). 
As a concrete implementation of their proposal they investigated statistics-changing maps induced by nonhomogeneous subspaces of Lie superalgebras closed under the superbrackets. This leads to the notion of {\it Volichenko algebras} which satisfy a condition known as metaabelianess (for any $X,Y,Z$ triple of operators, the identity
$ [X,[Y,Z]] = 0 $, which involves ordinary commutators, is satisfied).  Interestingly, they also introduced in \cite{lese2}
mixed brackets which interpolate ordinary commutators and anticommutators (they could be expressed in terms of mixing angles, but in their construction Leites-Serganova did not determine the angles).\par
~\par
The matrices induced by the realization of the braided tensor products in terms of (via the intertwining operators) the ordinary tensor products, naturally induce a metasymmetry following the Leites-Serganova terminology. In the simplest case, the algebra of $2$-particle creation/annihilation operators defines a mixed-bracket generalization
of the $2$-oscillator Heisenberg-Lie algebra.  The mixed-bracket $(.,.)$ can be defined as
\bea\label{volimixedbracket}
(X,Y)_{\vartheta_{XY}} &:=& i \sin(\vartheta_{XY}) \cdot [X,Y] + \cos(\vartheta_{XY})\cdot \{X,Y\},
\eea
The $5$ generators ($G_{\pm1}, G_{\pm 2}, G_0$) $2$-oscillator algebra has the only nonvanishing brackets given by
\bea
&(G_{\pm 1}, G_{\mp 1}) =  (G_{\pm 2}, G_{\mp 2}) = G_0.&
\eea
The mixed-bracket formulas, with the  explicit insertion of the $\vartheta_{IJ}$ angle dependence, are
\bea\label{anticommut}
&(G_{+1},G_{-1})_0=(G_{-1},G_{+1})_0= G_0, \qquad (G_{+1},G_{+1})_0=(G_{-1},G_{-1})_0 =0,&\nonumber\\
&(G_{+2},G_{-2})_0=(G_{-2},G_{+2})_0= G_0, \qquad (G_{+2},G_{+2})_0=(G_{-2},G_{-2})_0=0,&
\eea
together with 
\bea\label{genuinemixed}
&(G_{+1},G_{+2})_{+\frac{s+2}{2s}\pi}=(G_{+2},G_{+1})_{-\frac{s+2}{2s}\pi}= 0, &\nonumber\\
&(G_{+1},G_{-2})_{-\frac{s+2}{2s}\pi}=(G_{-2},G_{+1})_{+\frac{s+2}{2s}\pi}= 0, &\nonumber\\
&(G_{-1},G_{+2})_{-\frac{s+2}{2s}\pi}=(G_{+2},G_{-1})_{+\frac{s+2}{2s}\pi}= 0, &\nonumber\\
&(G_{-1},G_{-2})_{+\frac{s+2}{2s}\pi}=(G_{-2},G_{-1})_{-\frac{s+2}{2s}\pi}= 0. &
\eea

At $s=2$ these brackets define an ordinary, $2$ fermionic oscillators, Heisenberg-Lie algebra. 
For any given $s=3,4,5,\ldots$, they are a mixed bracket generalization of the Heisenberg-Lie algebra which encodes the braid statistics of level $s$. 
Since $G_0$ is a central element, these level$-s$ mixed-bracket algebras not only satisfy generalized Jacobi identities; they also satisfy a ``metaabelianess condition with respect to the mixed brackets\cc :
\bea
(G_I,(G_J,G_K)) &=& 0 \qquad {\textrm{for any ~~$I,J,K=0,\pm 1,\pm 2$
}}.
\eea
The ordinary metaabelianess condition is not satisfied by $G_{\pm1}, G_{\pm 2}, G_0$ as it can be easily checked. A possible explanation is that the Leites-Serganova construction is {\it classical}. Their notion of metaspace is applied to classical supergeometry and classical Lie superalgebras. On the other hand, the construction here presented is related to quantum groups at roots of unity. Quite naturally, a new question arises: is it possible to introduce a notion of ``quantum metaspace\cc~which seems has not yet being considered in the mathematical literature? The very recent results here presented suggest that this could be the case.\par
~\par
In the conclusion I make a connection with the \zzg parafermionic oscillators discussed in Section {\bf 3}. 
In the $s\rightarrow \infty$ limit (the untruncated case) the mixed brackets produce ordinary commutators and anticommutators (that is, there are no mixed terms since either the $\sin$ function or the $\cos$ function entering the right hand side is vanishing). Despite being ordinary commutators/anticommutators,
the brackets are arranged in a peculiar way. Indeed, they define a ${\mathbb Z}_2^2$-graded parafermionic oscillator
algebra.
Both sets of $\{G_0,G_{\pm 1}\}$ and $\{G_0,G_{\pm 2}\}$ subalgebras close a fermionic oscillator algebra:
\bea\label{paraf1}
\relax &[G_0, G_{\pm 1}] =0, \qquad \{G_{\pm 1}, G_{\pm 1} \}=0, \qquad \{G_{+1}, G_{-1}\} = G_0,&\nonumber\\
\relax &[G_0, G_{\pm 2}] =0, \qquad \{G_{\pm 2}, G_{\pm 2} \}=0, \qquad \{G_{+2}, G_{-2}\} = G_0.&
\eea
On the other hand the brackets recovered fronm one $G_{\pm 1}$ and one $G_{\pm 2}$ generator are given by ordinary commutators:
\bea\label{paraf2}
\relax [G_{\pm 1}, G_{\pm 2}] =0, && [G_{\pm 1}, G_{\mp 2}]=0.
\eea
The (\ref{paraf1},\ref{paraf2}) brackets define the ${\mathbb Z}_2^2$-graded parafermionic oscillators algebra.\par~\par

~\par
~
\\ {\Large{\bf Acknowledgments}}
{}~\par{}~\\
Along the years I have profited of the constant discussions and clarifications from my long-term collaborators N. Aizawa and Z. Kuznetsova. The quantization of braided Majorana qubits and related mathematical structures profited of important insights from D. Leites, V. Serganova and N. Reshetikhin.
 This work was supported by CNPq (PQ grant 308846/2021-4).

\end{document}